\documentclass[final,3p,times]{elsarticle}

\usepackage{amssymb}

\usepackage{siunitx}
\usepackage{chemformula}
\usepackage{comment}
\usepackage{soul}

\usepackage[font=small,labelfont=bf,labelsep=colon]{caption}
\begin{document}

\begin{frontmatter}



\title{Lift-off free catalyst for Metal Assisted Chemical Etching of silicon in vapour phase }

\author[inst1]{Hanna Ohlin}

\affiliation[inst1]{organization={KTH Royal Institute of Technology, Department of Applied Physics, Albanova University Center},
            postcode={106 91},
            city={Stockholm},
            country={Sweden}}

\affiliation[inst2]{organization={Department of Physics, University of Basel},
            country={Switzerland}}

\affiliation[inst3]{organization={Swiss Nanoscience Institute, University of Basel},
            country={Switzerland}}

\affiliation[inst4]{organization={Paul Scherrer Institut},
            postcode={5235},
            city={Villigen PSI},
            country={Switzerland}} 

\affiliation[inst5]{organization={Institute for Biomedical Engineering, University and ETH Zurich},
            postcode={8092},
            city={Zurich},
            country={Switzerland}}

\author[inst2,inst3,inst4]{Bryan Benz}
\author[inst4,inst5]{Lucia Romano}
\author[inst1]{Ulrich Vogt}

\begin{abstract}
Metal-assisted chemical etching of silicon is a promising method for fabricating nanostructures with a high aspect ratio. To define a pattern for the catalyst, lift-off processes are commonly used. The lift-off step however is often a process bottle neck due to low yield, especially for smaller structures. To bypass the lift-off process, other methods such as electroplating can be utilized. In this paper, we suggest an electroplated bi-layer catalyst for vapour phase metal-assisted chemical etching as an alternative to the commonly utilised lift-off process. Samples were successfully etched in vapour, and resulting structures had feature sizes down to 10 nm.

\end{abstract}



\begin{keyword}
MACE \sep MacEtch \sep metal assisted chemical etching \sep zone plate \sep nanostructures
\end{keyword}

\end{frontmatter}


\section{Introduction}
\label{sec:introduction}

Metal assisted chemical etching (MACE) is a method shown to be efficient for etching silicon micro- and nanostructures, and has since its advent been adapted for different kinds of devices \cite{Huang2011}. In the field of X-ray optics, structures with small feature sizes and high aspect ratios are needed to facilitate devices with best possible performance, e.g., zone plates or gratings. MACE as a method has shown potential in fabricating structures with such properties \cite{Chang2014,Akan2018,Li2020,Romano2020-2}. This can be done through different fabrication paths, where catalyst layers and etchant mediums are among the tunable variables \cite{Romano2017,Akan2021,Romano2020}. For focusing optics such as X-ray zone plates, the smaller the outermost zone of the structure is, the better the focus will be. This demands an increased precision when manufacturing the structures. To achieve this, the MACE process needs to be optimised \cite{Akan2018}.

To achieve a good result from a MACE process, steps need to be taken to facilitate a good catalyst layer preparation. A commonly used process for pattern definition is lift-off, following lithography and metalisation through some manner of evaporation. Whilst it is a tried and tested method, lift-off processing is not without challenges of its own, as both yield and outcome can vary greatly \cite{Ohlin2023}. Pattern definition of this kind, when desired feature size are at nanoscale, the pattern fidelity comes with non-trivial processes that might have a low throughput yield. This gives rise to the question whether other methods can be used to assemble the catalyst layer, rendering lift-off dispensable in processes that require small features.

In order to target high aspect ratio structures with features size in the range of 100 nm or even below, the stability of the etched lamellas in wet environment become relevant. MACE in vapor phase removes the stiction due to the liquid drying process after the etching step. Therefore, this process could be attractive for nanofabrication of zone plates or other small features high aspect ratio structures.

In this paper, we investigate a novel fabrication path for nanostructures made by metal assisted chemical etching in vapour. By electroplating a platinum layer over a gold seed layer, a lift-off-free pattern definition process is achieved which results in a thicker bi-layer Au/Pt catalyst. Residual gold seed layer is stripped away utilising RIE, before samples are etched through MACE in vapour phase.

\begin{figure*}[!h]
    \centering
    \includegraphics[height=6.2cm]{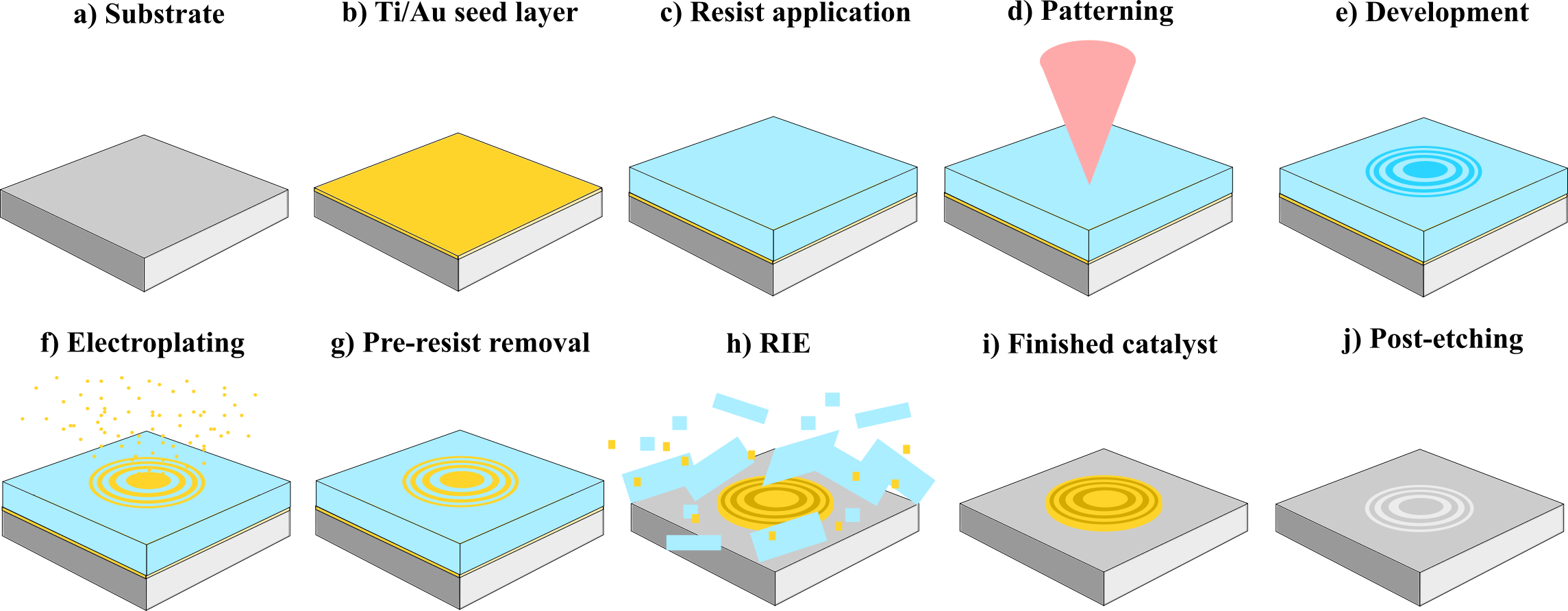}
    \caption{An overview of the in-going process steps, from clean sample to printed, plated structure. The top row shows how the sample goes from a) raw substrate, b) seed layer metalisation of Ti and Au, c) application of resist, d) e-beam lithograpy and e) development. In the bottom row, f) and g) shows electroplating during and after the process, h) displays the RIE-process that strips resist and residual seed layer, to reveal i) the finished catalyst. Lastly, j) shows the sample post-etch.}
    \label{fig:process}
\end{figure*}

\section{Materials and method}
\label{sec:method}

Samples for the etching experiments were prepared in a step-by-step process as follows.

\subsection{Sample preparation}
Silicon of both N (phosphor doped, 1-5 Ohm/cm resistivity) and P-type (boron doped, 1-5 Ohm/cm resistivity) were cleaved to 1 cm x 1 cm large pieces, before sonicated in isopropanol for 5 minutes to remove any silicon residuals from the dicing (Fig.\ref{fig:process}a). Samples were then cleaned through a RIE-process. Through electron beam evaporation, a sticking layer of 3 nm Ti was added, followed by a 10 nm Au seeding layer to facilitate any following electroplating steps (Fig.\ref{fig:process}b). A 90 nm layer of AR-P 6200.09, also known as CSAR62, (AllResist GmbH) was added by spincoating before pattern definition (Fig.\ref{fig:process}c).

\subsection{Patterning}
To pattern each sample, electron beam lithograpy was performed using a RAITH VOYAGER 50 kV system at a 120 µC starting dose with radial dose increase for proximity effect correction. Zone plate patterns were used as test structures, as these offer a wide range of feature sizes. These patterns have previously been used for similar experiments \cite{Ohlin2023-2}. Each sample was printed with 10 zone plate structures, where the smallest outermost feature size was 30 nm. The intended line-to-space-ratio was 1:1. Samples were developed in amyl acetate for 60 seconds, before development was stopped by a 10 s submerge in IPA. A final rinse in pentane for 20 seconds concluded the development process. This resulted in a mask for the following electroplating step, with an open resist pattern revealing the gold seed layer below (Fig.\ref{fig:process}e).

\subsection{Metallisation}
Samples were electroplated in a H\(_2\)SO\(_4\)-based platinum electroplating bath (JE18, Jentner Plating Technology GmbH) at a fixed voltage of 2 V for 17 seconds (Fig.\ref{fig:process}f-g). No reference electrode or area was used. The setup has previously been described elsewhere \cite{Ohlin2022}. The current was left free-floating so to ensure a constant voltage. Resist was removed through sonicating the plated samples in the remover solution AR 600-91 (Allresist GmbH) for 20 minutes. A finalising RIE (Oxford Instruments, Plasmalab 80 Plus) step of 15 minutes at 20 sccm O\(_2\), 250 W ICP RF was conducted to assure a clean surface and that all the resist had been properly removed before etching (Fig.\ref{fig:process}h). It is important to note that the gold seed layer outside the electroplated structures was also stripped in the same RIE process to minimize any etching outside the bilayer catalyst structure.

\subsection{Metal assisted chemical etching}
The MACE experiment was realized in a vapor-HF tool as reported previously\cite{Romano2020}. The liquid solution was prepared by mixing different volumes of water diluted HF (49 wt\(\%\)) and deionized water (18 MOhm$\cdot$cm), and the total liquid volume of etchant was fixed to 200 ml. The sample holder was warmed at temperatures higher than 35\(^\circ\)C during etching to prevent the liquid condense, and the distance d between the sample surface and the liquid level can be varied between 0.5 and 2 cm. We fixed the etching time to 15 min. The etching parameters (HF composition, holder temperature and distance from the liquid level) have been varied assessing the etching quality of the patterns via SEM inspection. Table~\ref{tbl:etching_conditions}. reports the sets of used conditions. The HF concentration and the distance were used to vary the etching rate, with the fastest etching at highest distance and highest HF concentration. The MACE reaction occurs in a vapour of HF and air, so the oxidant is the O\(_2\) gas component of air. Pt has a very high reactivity to reduce O\(_2\). In contrast, previous experiments with Au catalyst (not shown) indicated an etching rate in vapour MACE of at least 2 orders of magnitude lower with respect to Pt.

\begin{figure}
    \centering
    \includegraphics[height=3.6cm]{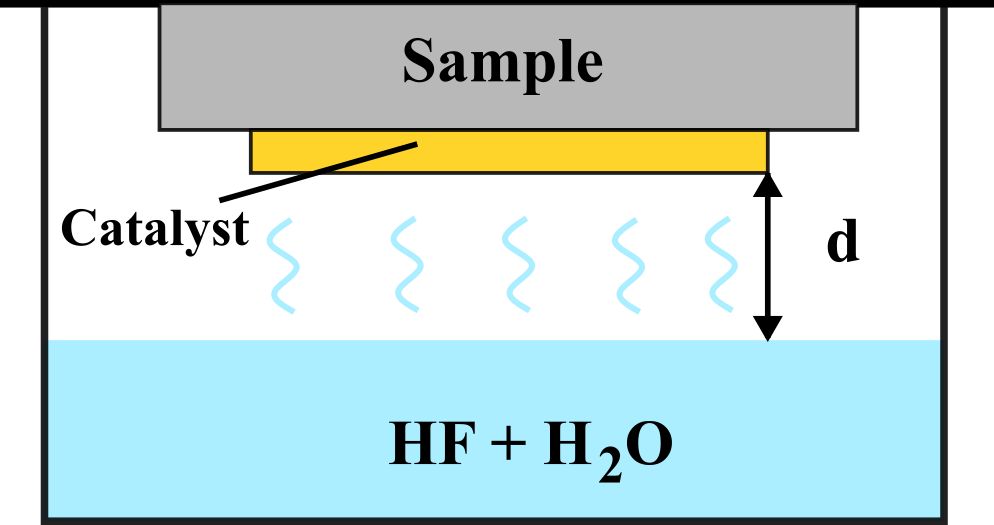}
    \caption{A visualisation of the setup of a vapour phase MACE experiment. The sample is suspended above the etchant solution that is allowed to evaporate up towards the sample, etching it. The distance between the sample and etchant is denoted by d, and can be varied.}
    \label{fig:vsetup}
\end{figure}

Metal assisted chemical etching was conducted in vapour phase, and reaction conditions were adapted from previous experiments \cite{Shi2023}. A visualisation of a vapour phase set up can be seen in Fig.\ref{fig:vsetup}.

\begin{table}[h]
\small
  \caption{\ Etching conditions for sequential etch experiments performed. T denotes the temperature of the sample holder, HF:Di the volume fraction with respect to the added DI water and distance d of the sample from the liquid level in centimeters.}
  \label{tbl:etching_conditions}
  \begin{tabular*}{0.48\textwidth}{@{\extracolsep{\fill}}llll}
    \hline
    Exp. No & T(C) & HF:DI ratio  & d (cm) \\
    \hline
    1 & 50 & 1:1 & 1.3 \\
    2 & 35 & 1:1 & 1.3 \\
    3 & 35 & 1:1 & 2 \\
    4 & 35 & 1:0 & 2 \\
    5 & 50 & 1:0 & 0.5 \\
    \hline
  \end{tabular*}
\end{table}

Reaction conditions were continuously adapted to find a functioning interval for the etching process, and results were evaluated between each change to the reaction conditions.

\subsection{Analysis}
Samples were images using scanning electron microscopy (SEM). Both a Zeiss Supra VP55 and a Zeiss Leo 1530 were used. Height measurements of plated structures were done utilising profilometry (KLA Tencor P-15, Milipitas, CA, USA) with 0.5 mg stylus force.


\section{Results}
\label{sec:results}

Samples have been imaged both before and after etching, and for the sake of adapting the reaction conditions continuously during the experiment series.

\subsection{Catalyst layer}
\begin{figure*}[!h]
\centering
\includegraphics[height=3.6cm]{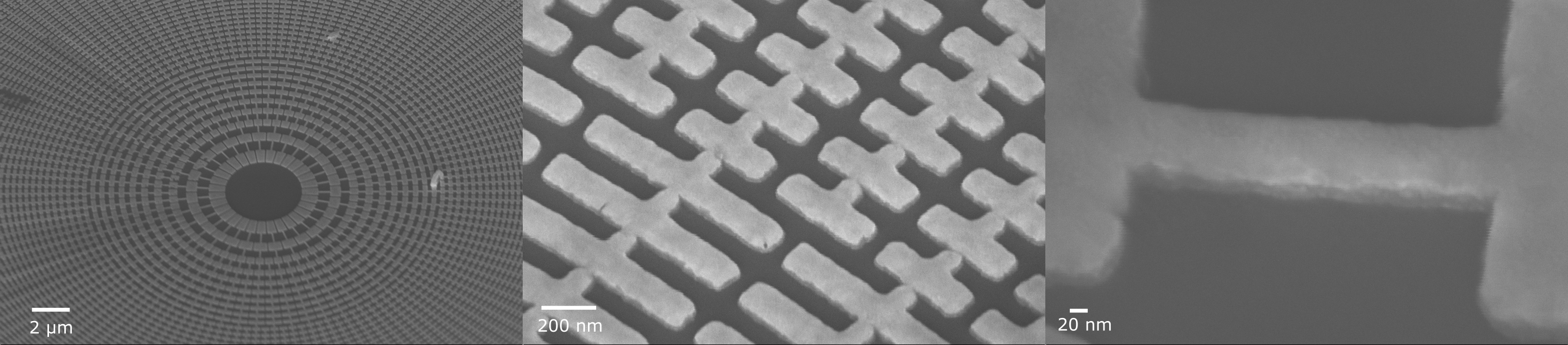}
\caption{The catalyst layer after resist removal imaged under 40\(^\circ\) tilt. From left to right, details of a plated zone plate pattern is shown in increasing magnification. The leftmost picture shows the center of the structure, where the increasingly thinning zones radially can be seen. The middle shows a detail from the middle zones of the zone plate, with clear fins on longitudinal spines. Lastly, a close-up of a part of a spine is shown. The layer has a 3-dimensional nature to it, and is thick enough to be imaged.}
   \label{fig:CL}
\end{figure*}

The catalyst layer was studied both in plain view and under tilt. An example is shown in figure~\ref{fig:CL}.

To the left, a center of the zone plate is shown. With increasing magnification, the structures are shown in more in-going detail. The Au/Pt-bilayer is thicker than catalyst layers that have been made for similar structures in the past \cite{Ohlin2023-2}. Furthermore, the catalyst layer shows to be relatively smooth, with no apparent anomalies or coarseness to it. The bi-layer in figure~\ref{fig:CL} is 60 nm thick as measured by mechanical profilometry. The catalyst thickness has been measured for all the samples and varies between 45-60 nm, and is uniform across each sample.

\subsection{Etch results}

\begin{figure*}[h]
    \centering
    \includegraphics[height=5.2cm]{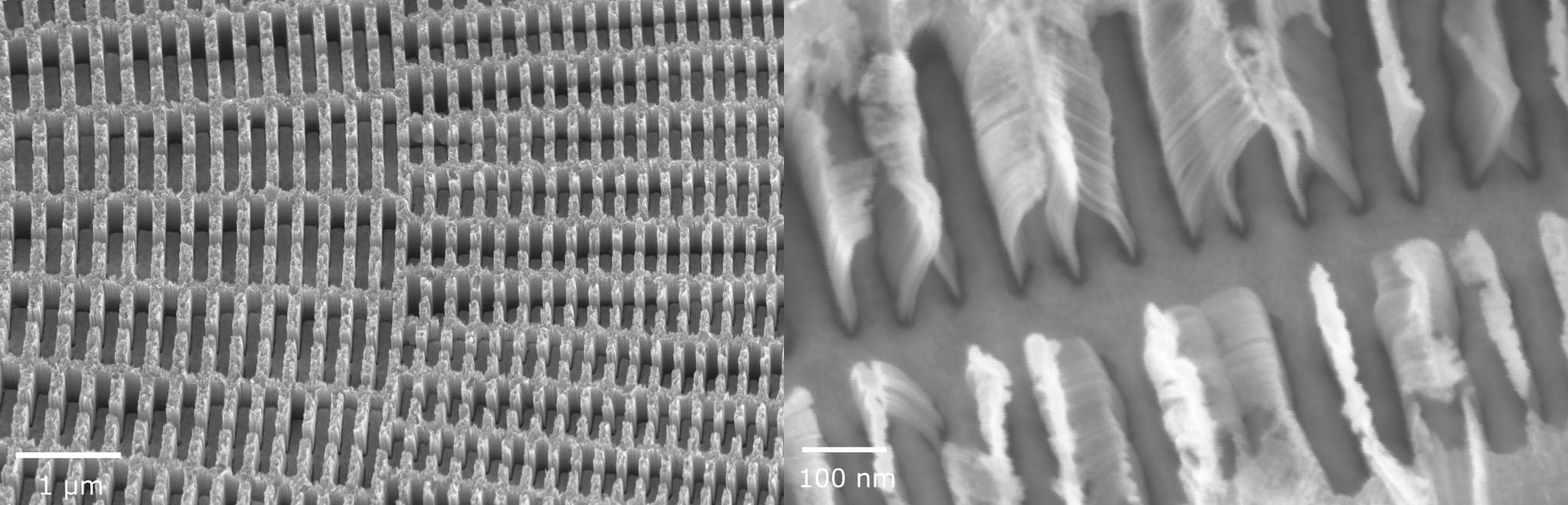}
    \caption{Tilted imaging of intermediate zones under 40\(^\circ\) tilt. Whilst pattern integrity is retained, some porosity can be seen. To the right, outermost zones are imaged in high resolution, revealing lamellae with a thickness in the 10 nanometer-range.}
    \label{fig:tilt}
\end{figure*}

After etching, the samples were then imaged anew under tilt for the sake of observing the etch results. An example is shown in figure~\ref{fig:tilt}.

Despite a somewhat porous etch result, the plated catalyst layer has etched the silicon in such a fashion that the intended pattern is obtained. The detailed picture also shows that not only thicker structures remain reasonably intact, but also thinner structures. The smallest features fabricated were in the 10 nm range and around 500 nm tall, although collapsed from lack of structural integrity. These were fabricated in experiment no 5 in table~\ref{tbl:etching_conditions}. Judging by the catalyst layer, they seem to not have eroded from the sides, even if, as previously mentioned, the top of the structures have been rendered porous. The resulting porosity is likely due to the etching conditions needing further optimization than the attempted conditions for this experiment. 

\subsection{Cross-sections}
\begin{figure*}[h]
    \centering
    \includegraphics[height=7cm]{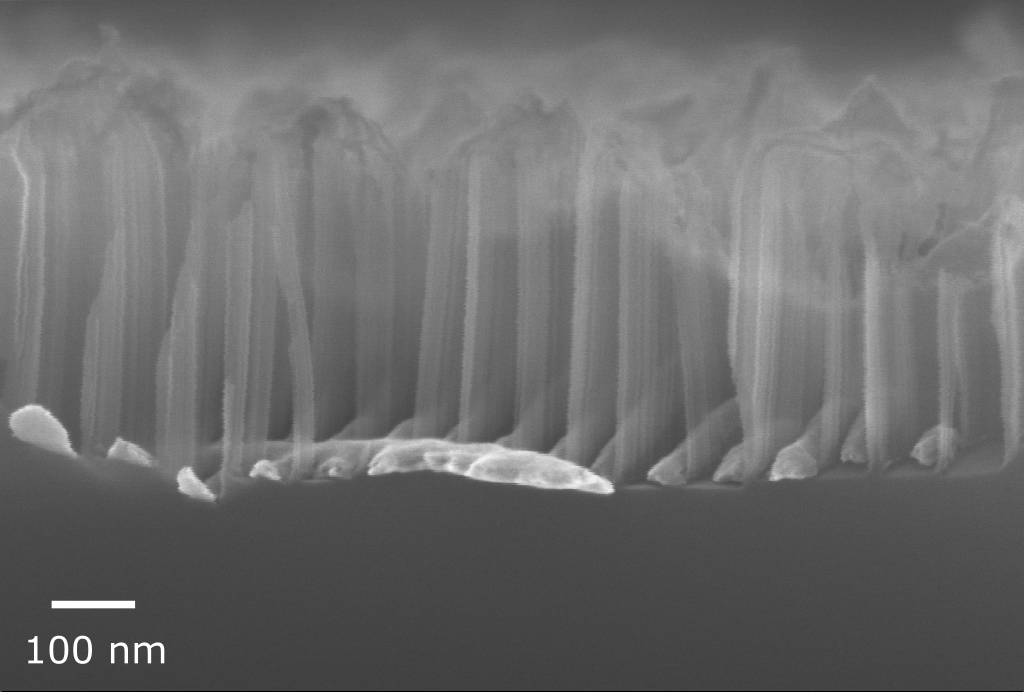}
    \caption{Detail picture of a cleaved sample imaged on the cross-section. In the very bottom of the lamellar structure, the catalyst layers are visible, both fins and spines intact. The very thin silicon lamellae are intact, yet eroded in the top of the structure.}
    \label{fig:CS}
\end{figure*}

Samples were diced and imaged in cross section at a 90 degree angle. An example of a cross-section is shown in figure~\ref{fig:CS}.

Studying the cross sections in figure~\ref{fig:CS} fabricated in experiment no 5 in table~\ref{tbl:etching_conditions} reveals that the catalyst layer indeed etches, and that it seems to do so in a vertical fashion. There is still a tangible issue with errant etching in the top of the structures, but this is not shown in the same way in the depths, where structures remain thin and intact. At the bottom of the etched structures, the catalyst layer can clearly be seen in lighter shades at the bottom of the silicon trenches. Comparing to the images in figure~\ref{fig:CL}, it does not seem like the catalyst layer has eroded, but shape and structure has been retained all the way down to the bottom of the etched trenches.

\section{Discussion}
\label{sec:discussion}
Discussing the etching results from this novel catalyst, a few things can be noted.

\subsection{Porosity and depth}
As visible in the imaging results, in both the tilted images and the cross section images, it is clear that there is an overwhelming amount of porosity over the surface next to the etched structures. That MACE processing of silicon renders the substrate porous is a known issue \cite{Li2000,Ohlin2023-2,Geyer2015}. Thus, the actual etch depth is hard to determine, as there is no judging whether the surface of the sample has etched away or if it still remains in part as a starting point for a depth measurement. What can be seen clearly is however that there is a degree of verticality to the structure that is high enough to preserve even very small features. The porosity certainly complicates matters, but this is potentially something that could be resolved with more testing and more optimisation of the etch conditions. Etching with platinum catalysts has been done before in a successful fashion \cite{Romano2020}, and the proven verticality and retention of small features allude to there being process parameters that would fit this catalyst layer with enough optimisation.

\subsection{Catalyst attributes}

The resist layer was exposed in the shape of a zone plate with a 1:1 line-to-space ratio, where the outermost zones were 30 nm. The same exposure parameters have been previously used in similar experiments, but for lift-off with evaporated gold and liquid phase MACE \cite{Ohlin2023-2}. As the catalyst layer here is electroplated, there might be a shift in the line-to-space ratio as the platinum grows not only upwards, but potentially slightly sideways due to a resist undercut. The widening of the zones can be seen in figure \ref{fig:CL}.

The catalyst layer is made up of three components, where two layers - gold and platinum - are assumed to drive the reaction. This being said, the third titanium sticking layer might play a part in the reaction as well, as the flow of residual and waste products from the etch reaction is not as fluid in vapour as it is in liquid phase. As gold does not readily adheres to silicon, the titanium sticking layer is necessary to ensure adhesion of the catalyst to the substrate. What exact impact the thin Ti-layer (3 nm) has on the reaction remains a point which should be investigated further. Previous literature  has reported a sort of blocking action from Ti layer (5-10 nm) between Au catalyst layer and Si substrate \cite{Kong2016}.

\subsection{Reaction conditions}

The initial conditions were adapted from previously made experiments that have shown good results for platinum catalysts \cite{Romano2020}. What is interesting is that despite platinum making up for the bulk of the bilayer catalyst, it does not seem to behave like previously studied evaporated platinum layers. Bilayer catalysts have been studied in the past \cite{Kim2011,Wu2016}, but either with different metals or with different feature sizes and etch mediums. Discerning what precisely causes the unwanted porosity is challenging, and more experimentation and optimisation is needed to limit the influence of it on the structures. Similarly, the reaction rate is not alike previously conducted experiments with similar structures \cite{Romano2020,Ohlin2023-2}. This raises the question of how the charge carriers travel through the catalyst and how the catalyst truly interacts with the silicon layer during the process. The bottom gold layer is thick enough to separate the platinum and the silicon. At the same time, the reaction progresses in a more rapid fashion than what has previously been seen with just gold in vapour phase. To assume the platinum somehow drives the reaction rate is not unreasonable, the question is to what extent. 

Despite the large range of parameters used (see Table 1), the etching rate seems not affecting the etching quality. All the etched structures look very porous, contradicting the usual low porosity of MACE in vapor phase \cite{Hildreth2014}. This effect can be a result of the catalyst geometry, since the top layer of Pt is very thick so that the charge carriers can be directly injected into the etched Si lamellas from the sides of the Pt deposit. Moreover, the presence of Ti as sticking layer for Au deposition on Si substrate might play a role as additional interface for the charge carrier injection into the substrate, which might be slower than the lateral injection, resulting into a faster porosification of the Si lamellas. Usually, the Ti layer is etched off in liquid MACE at the very beginning of the process, while it is known that metal oxidation is slowed down in vapor-HF \cite{Witvrouw2000}. However, our preliminary experimental results demonstrate that very thin lamellas (sub 10 nm), as shown in figure \ref{fig:tilt}, can be realized in a patterned structure with a multi layer catalyst and without lift-off, demonstrating the capability of pattern transfer and the potential resolution at nanoscale of the technique.

\section{Conclusion}
\label{sec:conclusion}

We have shown that it is possible to etch samples with a bi-layer catalyst through metal assisted chemical etching in vapour phase. Whilst the process needs further optimization, we have proven that plated catalyst bi-layers work for vapour MACE, and that they under certain circumstances may etch even complicated, very small and dense structures in a vertical fashion. Further investigations are necessary to optimise the etching conditions, as well as the composition of the bi-layer and its attributes.

\section{Acknowledgements}
This research was funded by the Swedish Research Council grant number 2018-04237 and 2019-06104, as well as PHRT-TT Project Nr. 2022-572 INTIMACY, SwissLOS Lottery Fund of Kanton Aargau, SNI PhDSchool Project Nr. P2205 MAGNET. The authors wish to acknowledge the clean room facilities of PSI to which we express our gratitude, and Prof. Marco Stampanoni for PSI hosting. The authors also wish to acknowledge the contribution and support of Markus Ek at the Department for Chemistry at Uppsala University for aid in SEM imaging and analysis, and to Hazal Batili at the Department of Applied Physics at KTH Royal Institute of Technology for aid with MACE experimentation and lab support.

\section{Author Contribution}
H.O. conceived the presented idea, performed the patterning, carried out the characterizations; B.B. carried out the MACE experiments and part of the SEM characterization; L.R. supervised the project at PSI, U.V. supervised the project at KTH. All authors provided critical feedback and helped shape the research, analysis and manuscript.

 \bibliographystyle{elsarticle-num} 
 \bibliography{MACE_Refs}

\end{document}